\documentclass[a4paper,10pt]{article}
\usepackage[affil-it]{authblk}
\usepackage{epsfig}
\usepackage{graphicx}
\usepackage{color} 
\usepackage{amsfonts}
\usepackage{amsmath}
\usepackage{fancyhdr}
\RequirePackage{fix-cm}
\usepackage[paperheight=25cm,left=2.2cm,right=2.2cm,top=2.0cm, bottom=2.5cm]{geometry}

\newcommand{\ct}{\cite}
\newcommand{\bi}{\bibitem}
\newcommand{\be}{\begin{equation}}
\newcommand{\ee}{\end{equation}}
\newcommand{\ba}{\begin{eqnarray}}
\newcommand{\ea}{\end{eqnarray}}

\fancypagestyle{plain}{%
   \fancyhead[L]{\fbox{C V Raman special issue (Indian Journal of Physics), DOI: 10.1007/s12648-014-0483-9}}
   
}
\begin{document}
\title{Quantum Annealing search of Ising spin glass ground state(s) with tunable transverse $\&$ longitudinal fields}                                                                              

\author{A Rajak%
  \thanks{Electronic address: \texttt{atanu.rajak@saha.ac.in}}}
\author{B K Chakrabarti%
  \thanks{Electronic address: \texttt{bikask.chakrabarti@saha.ac.in}}}
\affil{Condensed Matter Physics Division \\ Saha Institute of Nuclear Physics \\ 1/AF Bidhannagar, Kolkata 700 064, India.}

\maketitle
\begin{abstract}
\noindent Here we first discuss briefly the quantum annealing technique. We then study the quantum annealing of Sherrington-Kirkpatrick spin glass model 
with the tuning of both transverse and longitudinal fields. Both the fields are time-dependent and vanish adiabatically at the same time, starting from high values. 
We solve, for rather small systems, the time-dependent 
Schrodinger equation of the total Hamiltonian by employing a numerical technique. At the end of annealing we obtain the final state having high overlap with 
the exact ground state(s) of classical spin glass system (obtained independently). 
\end{abstract}
\textbf{$~~~~~~$Keywords:} Quantum annealing; Spin glass; Quantum tunneling
\vskip 0.2cm
\textbf{$~~$PACS Nos.:} 03.67.Lx; 75.10.Nr; 75.45.+j

\section{Introduction}
\label{I}
\begin{figure}
\centering
\includegraphics[height=2.6in]{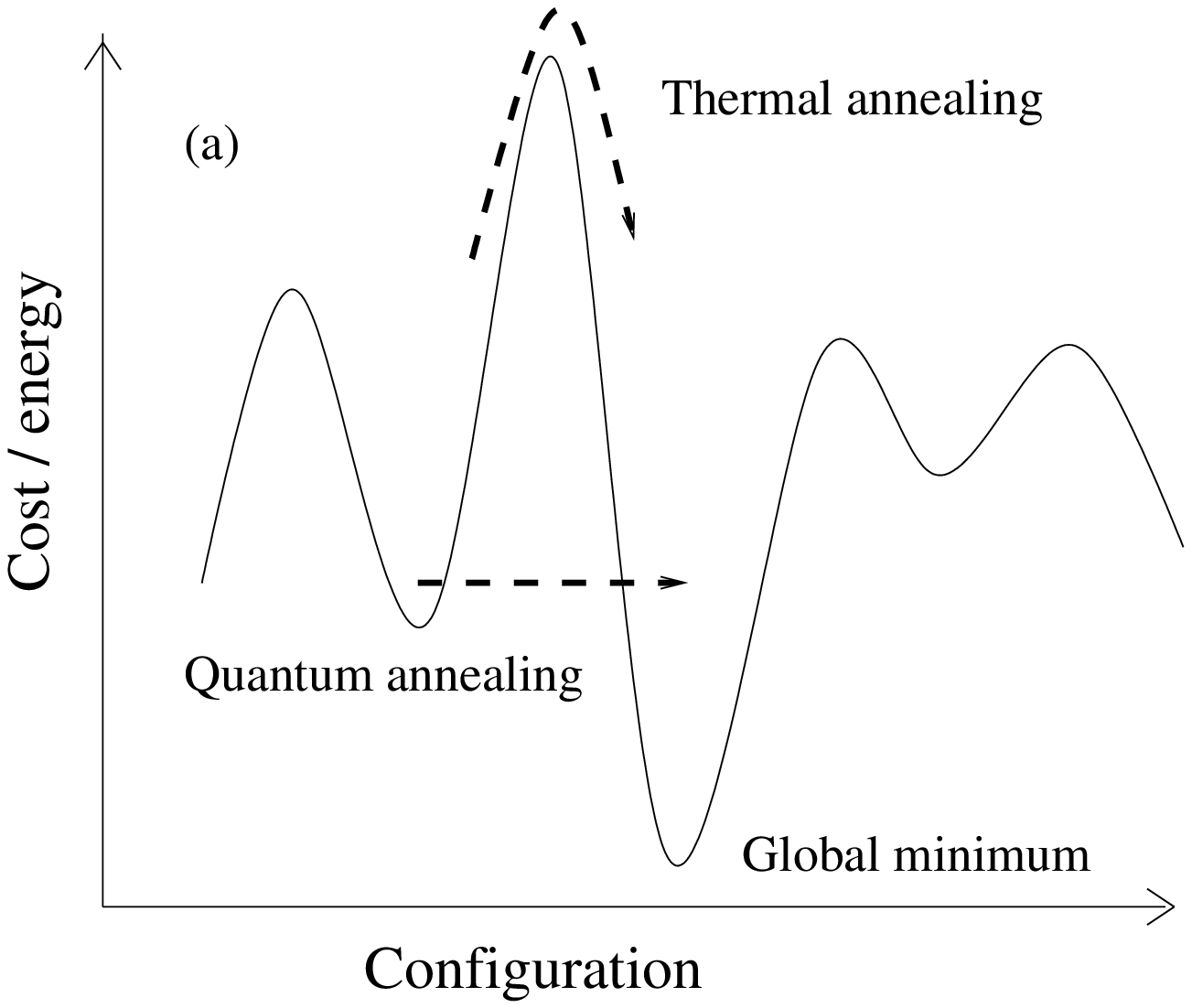}
\includegraphics[height=2.6in]{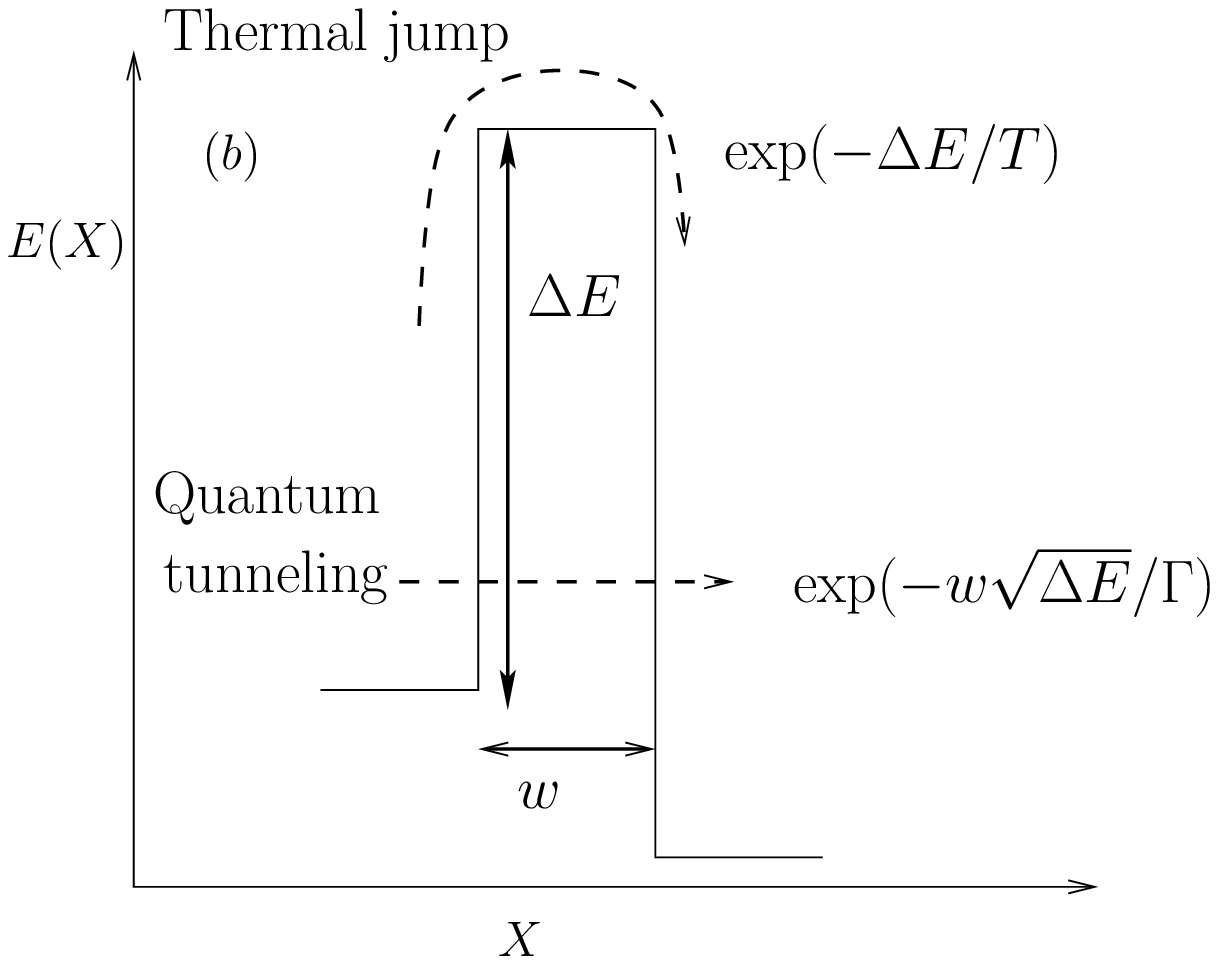}
\caption{ Variation of cost energy with different configurations of a computationally hard problem like in a spin glass. The figure $1(b)$ 
is a cartoon for figure $1(a)$. In classical annealing case, to reach global minima the system 
has to overcome a large barrier $\Delta E$ of $O(N)$, $N$ being the system size (escape probability $\sim \exp(-\Delta E/T)$ at temperature $T$). In quantum annealing case, system can tunnel through the barrier; 
if the barrier is narrow (tunneling probability $\sim \exp(-w\sqrt{\Delta E}/\Gamma)$ for tunneling fluctuation field $\Gamma$ and barrier width $w$).}
\label{fig_config}
\end{figure}

Optimization of cost function in a system with $N$ independent variables often becomes NP-hard, where the search time can not be bounded by any polynomial in $N$.
Such multivariable optimization problems can be represented by a search problem for the minima in a landscape with the effectiveness of the solution or cost plotted in the $y$-axis and the solution 
 (or configuration) number plotted in the $x$-axis. Typically, such cost function landscapes are very rugged and have
many local minima with one or degenerate global minima (see Fig.~\ref{fig_config}) in configuration space. So, search algorithms along a valley in such a landscape are not useful,
 because the system ends up in a local minimum of that valley which is not necessarily a global one. 
Kirkpatrick et al. \ct{kirkpatrick83} and Cerny \ct{cerny85} independently suggested a method called simulated annealing (SA) to solve such optimization problems. In this process, 
a stochastic algorithm is chosen with a tunable noise or fluctuation such that as the system gets into a local minimum (with cost $E_l$), the fluctuation permits for acceptance of higher 
cost value $E$ $(=E_l+\Delta E)$ solutions with a Boltzmann like probability $\exp(-\Delta E/T)$ and the system  is allowed to explore 
full configuration space. The global minimum (ground sate; may be degenerate) of the cost function is obtained as the fluctuation is properly tuned (with an annealing 
schedule) down to zero. 
Stated more explicitly, in this procedure initially one starts with an arbitrary configuration $C_i$ with cost function $E_i$ and 
go the configuration $C_f$ with energy $E_f$ following some stochastic rule. If $E_f$ is less than $E_i$ then one always accepts the change, otherwise accepts the 
change with finite probability $\exp(-\Delta E/T)$, where $\Delta E=E_f-E_i$. So, as the temperature decreases with time ($t$) to zero, hopefully the system attains a ground state 
configuration. The success of reaching of the true ground state depends on the annealing schedule $T(t)$ : If temperature is reduced too quickly the system may get  
localized in a local minimum and the obtained configuration is not that of a global minimum. In this regard, Geman and Geman have shown that if annealing schedule is 
taken as $T(t)\geq N/\ln t$, then the system reaches the ground state eventually \ct{geman84}. In this schedule 
a system must reach that with minimum energy, but it may take much longer time. In practice, good annealing results can be obtained through even a faster decrease of temperature. 

The simulated annealing (SA) is quite an useful technique to solve systems with many local minima in configuration space. But this technique does not give satisfactory 
results for a non-ergodic system like spin glass \ct{das04,das08}. In this systems there are many local minima and the minimum energy configurations are separated by the 
barrier heights of order of $N$. So, with thermal dynamics at any finite temperature, the full phase space is not covered by the system and the system may get trapped in local minima (for $N\rightarrow\infty$). Finite thermal fluctuations are 
unable to make the system ergodic (in the $N\rightarrow\infty$ limit) and the system can not go from one local minima to another. In such situations quantum annealing (QA) can be a good choice to overcome 
such effects and produce better results \ct{chandra11,bose13,ghosh13}. As argued by Ray et al. \ct{ray89}, the quantum tunneling (as indicated in Fig.~\ref{fig_config}) can help getting out of
the local minima and approach the optimum state (ground state) if the energy 
barrier is narrow (and the phases of the turning waves from different walls do not conspire to make the system localized). In QA process one considers a time-dependent 
Hamiltonian
\be
\mathcal{H}=\mathcal{H}_0+\Gamma(t)\mathcal{H_I},
\label{th}
\ee
where $\mathcal{H}_0$ is the classical time-independent Hamiltonian whose ground state is to be found. Here, $\mathcal{H'}(t)=\Gamma(t)\mathcal{H_I}$ is the tunneling term of 
the total Hamiltonian $\mathcal{H}$, which is added externally and does not commute with $\mathcal{H}_0$ ($[\mathcal{H}_0,\mathcal{H}_I]\neq 0$). $\Gamma(t)$ is the time-dependent tunneling strength between 
different classical configurations and plays similar role as temperature in SA.
 The additional part of the total Hamiltonian actually 
gives internal dynamics in the system and allows mixing of the different stationary states of $\mathcal{H}_0$. To get the full dynamics of the system we have to solve 
time-dependent Schordinger equation of the total Hamiltonian
\be
i\frac{\partial|\psi(t)\rangle}{\partial t}=\mathcal{H}(t)|\psi(t)\rangle,
\label{wf1}
\ee
where $|\psi(t)\rangle$ denotes the instantaneous wave function of the total Hamiltonian $\mathcal{H}$. The term $\mathcal{H'}(t)$ gives the quantum fluctuation 
in the system and helps the system to visit different eigenstates (configurations) of the (classical) Hamiltonian $\mathcal{H}_0$ by tunneling through the barriers. Initially 
the quantum fluctuation term ($\mathcal{H}'$) is kept very large compared to the classical part ($\mathcal{H}_0$) and the dynamics is governed mainly by the fluctuation term. The fluctuation 
strength tuned by tuning the parameter ($\Gamma(t)$) which decreases adiabatically from high value to zero as time $t\rightarrow \infty$. If we take initial state (state at $t=0$) as the ground state of the 
total Hamiltonian, then according to adiabatic theorem \ct{sarandy04} instantaneous quantum states will be ground states at that parameter values and at the end of the 
annealing the system will attain the ground state of the classical Hamiltonian with a satisfactory finite probability. This, we can expect even for the 
disordered frustrated system like the spin glass systems.

In most cases like above, we know the exact initial ground state for high value of tunneling term. However in some systems we may not know the precise ground state of initial Hamiltonian 
to start the annealing process. In such cases also one can employ quantum tunneling term to anneal the system. 
Das et al. \ct{das05} have illustrated such a semiclassical treatment for kinetically constrained system (KCS) where they have shown QA is faster than SA. 
KCS is original East model with a kinetic constraint defined as the $i$-th spin can not flip if the $(i-1)$-th spin is down.
If the dynamics of spins is governed by the quantum fluctuations then the spin flipping probabilities come from the tunneling probabilities through a potential barrier 
of a double-well whose boundaries are infinitely high. One can easily argue that a quantum particle can tunnel through a potential barrier of height 
$\Delta E$ and width $w$ with a rate, $R\propto \exp(-w\sqrt{\Delta E}/\Gamma)$, where $\Gamma$ is the strength of quantum fluctuation which decrease with time (see Fig.~\ref{fig_config}). 
On the other hand, classical annealing (SA) is guided by the Boltzmann factor $\exp(-\Delta E/T)$. One can see that in case of QA, tunneling rate decreases exponentially 
with a exponent value $\sqrt{\Delta E}$ (provided the barrier width does not scale with $\Delta E$), whereas in the SA it is $\Delta E$. This makes QA to be a 
better option to achieve global minima with respect to SA, in most cases.

Spin glasses are the systems which have random ferromagnetic and antiferromagnetic interactions \ct{bkc06} between spins at different lattice sites. This induces frustration in 
such systems where all the interactions can not be satisfied simultaneously. As a consequence, the ground state of spin glasses are get highly degenerate \ct{binder86}. These frustrations lead to a large number of local minima 
which are separated by energy barriers which are often system size ($N$) proportional: One needs to flip a finite fraction of $N$ spins to go from one minimum to another. 
As a result, finding the global minimum or ground state (may be degenerate) of such systems is a hard problem (shown Eq.(\ref{ham1}) to be NP-hard for the Sherrington and Kirkpatrick (SK) model \ct{sherrington75}). 
Starting with the pioneering paper by Kadowaki and Nishimori \ct{kadowaki98}, the effectiveness of QA in search for the ground state(s) in S-K model has been established. 
However, these studies mostly apply a small steady symmetry breaking (longitudinal) field on the system to break the degeneracies of the ground and excited states in 
the S-K model. However, in principle this induces some error. 
In this paper we present a QA study of the S-K model 
by tuning not only the transverse field but also a longitudinal field and reduce them both eventually to zero. We solve the Eq.(\ref{wf1}) of the S-K model Hamiltonian with time-dependent transverse and 
longitudinal fields which decreases with time.
Here we have shown that when both longitudinal and transverse fields decrease, starting from large initial values to zero, the annealed state is more likely to converge to the 
ground state. 
We have studied this here numerically for a rather small system size 
($N=8$ only) and demonstrate the advantage. We expect this technique to work more effectively for large system sizes. To the best of our knowledge, none of the 
earlier studies considered quantum annealing in presence of tunable symmetry breaking (longitudinal) field, along with that tunable quantum fluctuation of the tunneling 
field. We indicate here the advantages.

The plan of the paper is as follows: In Sec.\ref{II} we introduce the model system i.e., the S-K model with transverse and longitudinal fields and adopt two annealing schedules. 
In Sec.\ref{III} we discuss our numerical results and discuss the benefit obtained using the technique. Finally we conclude in Sec.\ref{IV}.

\section{Model}
\label{II}
We consider the S-K model of classical spin glass of $N$ Ising spins put in transverse and longitudinal fields, represented by the Hamiltonian
\be
\mathcal{H}(t)=-\sum_{i<j}^NJ_{ij}\sigma_i^z\sigma_j^z-\Gamma(t)\sum_i\sigma_i^x-h(t)\sum_i\sigma_i^z,
\label{ham1}
\ee
where $\sigma_i^z$ and $\sigma_i^x$ are the $z$ and $x$ component Pauli spin matrices respectively. $J_{ij}$ are the quenched random interactions between each pair of spins 
at different sites (long-range interaction). These random variables follow a Gaussian distribution of zero mean and variance $J^2/N$, given by
\be
\rho(J_{ij})=\Big(\frac{N}{2\pi J^2}\Big)^{1/2}\exp\Big(\frac{-NJ_{ij}^2}{2J^2}\Big).
\label{dist}
\ee
Here, $\Gamma(t)$ is the tunneling amplitude between different stationary states of classical spin glass system ($\mathcal{H}$ with $\Gamma=h=0$). $h(t)$ denotes the time-dependent longitudinal field 
which breaks degeneracy in the system. We can denote the classical spin glass part of the total Hamiltonian (in Eq.(\ref{ham1})) by $\mathcal{H}_0=-\sum_{i<j}^NJ_{ij}\sigma_i^z\sigma_j^z$.

We consider functional forms of $\Gamma(t)$ and $h(t)$ such that initially the parameters have large values and subsequently decrease to zero very slowly at large time. 
Finally the instantaneous state $|\psi(t)\rangle$ will be one of the 
ground states of classical spin glass ($\mathcal{H}_0$) if the process is adiabatic. In \ct{kadowaki98}, the authors have studied QA of S-K model by transverse field (with a specific annealing schedule) in presence of a small but constant longitudinal 
field to break the degeneracy of the system. In general the ground state of a spin glass system changes with even a small longitudinal field 
due to the frustrations coming from randomness of exchange interactions. Hence, there is quite a high probability to end up with a state which is not a true spin glass ground state. 
Here, we consider that
 the longitudinal field also varies with time and finally vanishes at large time, similar to the transverse field. So, we do not face here any such problem of changing the 
ground state of spin glass.
We take the overlap of instantaneous ground state at
each time (obtained from the solution of the Schordinger equation) with the ground state of $\mathcal{H}_0$ (the spin glass ground state). First, we take the annealing schedules of transverse and longitudinal fields as
$\Gamma(t)=\Gamma_0/\sqrt{t}$ and $h(t)=h_0/\sqrt{t}$ respectively, where where $\Gamma_0$ and $h_0$ are 
the initial value of the fields at $t=1$.
We then consider also the time-dependent form of these fields as $\Gamma(t)=\Gamma_0/t$ and $h(t)=h_0/t$. As the time $t\rightarrow\infty$ the longitudinal as well as transverse field will vanish. We solve numerically the time-dependent 
Schordinger equation (Eq.(\ref{wf1})) using the Hamiltonian (\ref{ham1}). 
 We first get the initial ground state of the system by exactly diagonalizing the Hamiltonian (\ref{ham1}) with initial parameter values. We then 
calculate a quantity which is called probability of staying the system at the ground state of classical spin glass at time $t$, is given by $P(t)=|\langle\psi_0|\psi(t)\rangle|^2$, where 
$|\psi_0\rangle$ is the ground state wave function of the Hamiltonian $\mathcal{H}_0$. In short, we repeat the same study as in \ct{kadowaki98}, but with an additional 
annealing of the longitudinal field, as mentioned earlier.

\section{Numerical results}
\label{III}
\begin{figure}
\centering
\includegraphics[height=2.6in]{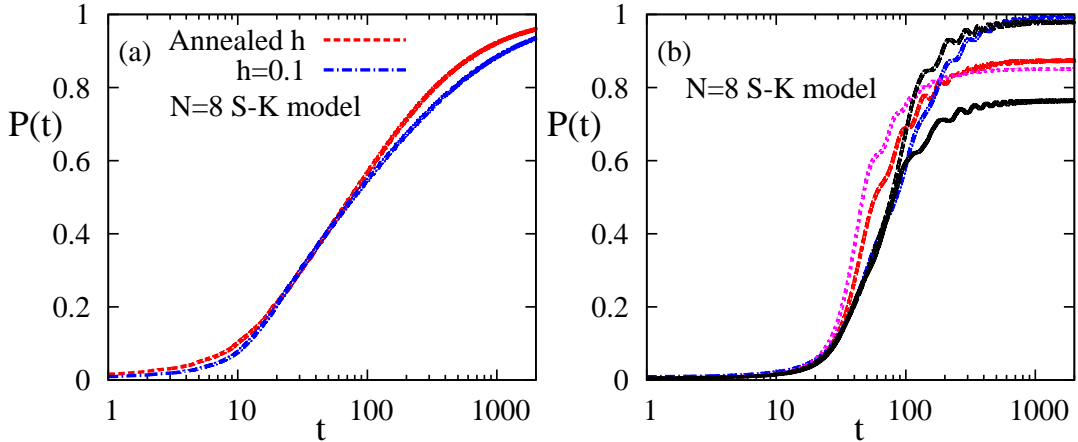}
\caption{(Color online) $(a)$ Overlap of instantaneous wave function with classical spin glass ground state with $\Gamma(t)=3/\sqrt{t}$ and $h(t)=-0.5/\sqrt{t}$. For comparison, we have shown the result for the same system with a fixed longitudinal 
field $h=0.1$ and same transverse annealing schedule.
$(b)$ Time variation of $P(t)$ for $5$ different sets of exchange interactions which have been taken from a fixed Gaussian distribution. 
Annealing with $\Gamma(t)=4/t$ and $h(t)=-1/t$.}
\label{fig_root}
\end{figure}

In this section, we show the numerical results using the annealing schedules of transverse and longitudinal fields as defined in Sec.\ref{II}.
First, we consider the annealing schedules as $\Gamma(t)=3/\sqrt{t}$ and $h(t)=h_0/\sqrt{t}$ where the value of $h_0$ may be positive or negative. Depending on the sign of $h_0$ the system will 
attain a ground state out of different degenerate ground states of spin glass.
 Here, we have considered $h_0=-0.5$ and the system converges to a (final) state 
 (which of course changes with the sign of $h_0$) at large time, and study its overlap ($P$) with the ground state of classical system with Hamiltonian $\mathcal{H}_0$
(with that particular configuration of $J_{ij}$).
The exchange interactions ($J_{ij}$) are taken from a Gaussian distribution (Eq.(\ref{dist})) with 
zero-mean (approximately) and standard deviation $0.5$. Fig.~\ref{fig_root}$(a)$ shows the time variation of the probability ($P(t)$) with annealing of $h$.
For comparisons, we also study the same with constant longitudinal field $h=0.1$ (as in \ct{kadowaki98}). 
The results shown here are 
for a typical set of exchange interactions. Some gain in the overlap is clearly seen for annealed $h(t)$ case.
In some isolated cases (with different set of exchange interactions), we have found little better results with annealing of $h$ than in the case of applying a small constant longitudinal field. 

We have also shown the numerical results (see Fig.~\ref{fig_root}$(b)$) for the annealing schedules $\Gamma(t)=4/t$ and $h(t)=-1/t$ with different realization of exchange 
interactions. One can see the overlap values (for the final annealed state with the exact ground state of $\mathcal{H}_0$) are quite good.
 We have found out that if we adjust the values of $\Gamma_0$ and $h_0$, the overlaps can 
increase still in some cases.

\section{Conclusions}
\label{IV}
After a brief introduction to quantum annealing (see e.g., \ct{ghosh13,bapst13,suzuki13} for some recent reviews), we
 have proposed here an annealing scheme to reach the exact ground state of disordered spin systems with a satisfactory probability by tuning of both transverse and longitudinal fields. Here, we have applied the scheme 
on a small-size ($N=8$) long-range interacting spin glass system. We have observed that applying even a small symmetry breaking (longitudinal) field on a S-K model (as done in usual earlier studies)
distorts the ground state(s) and induces significant errors (often as high as more than $50\%$). 
 Here, initially both the transverse and longitudinal field values are kept large and then both reduced to zero adiabatically as $t\rightarrow\infty$ 
(following identical schedule, either as proportional to $1/\sqrt{t}$ or $1/t$). 
In the process, the final state obtained has increased overlap with the exact ground state of classical spin glass (or $\mathcal{H}_0$). 
The obtained advantages (see Figs. \ref{fig_root}$(a)$,\ref{fig_root}$(b)$) are expected to be stable and and valid for larger system sizes (detailed and extended study results 
will be presented elsewhere).

\end{document}